\documentclass[11pt,a4paper]{article}
\pdfoutput=1
\usepackage{jheppub}
\usepackage{multirow, graphicx,amssymb,url,mathrsfs,amsmath}
\usepackage{wrapfig,boxedminipage,subfigure,epsfig}
\usepackage{amsxtra,amstext,latexsym,dsfont,amsfonts}
\usepackage{color}
\usepackage[dvipsnames]{xcolor}
\usepackage{float}
\usepackage{slashed}

\newcommand{\dd}{\mathrm{d}}

\title{Reflected Entropy and Entanglement Wedge Cross Section with the First Order Correction}

\author{Hyun-Sik Jeong,}
\author{Keun-Young Kim,}
\author{Mitsuhiro Nishida}

\emailAdd{hyunsik@gist.ac.kr}
\emailAdd{fortoe@gist.ac.kr}
\emailAdd{mnishida@gist.ac.kr}

\affiliation{ School of Physics and Chemistry, Gwangju Institute of Science and Technology, \\
123 Cheomdan-gwagiro, Gwangju 61005, Korea}

\abstract{We study the holographic duality between the reflected entropy and the entanglement wedge cross section with the first order correction. In the field theory side, we consider the reflected entropy for $\rho_{AB}^m$, where $\rho_{AB}$ is the reduced density matrix for two intervals in the ground state. The reflected entropy in the 2d holographic conformal field theories is computed perturbatively up to the first order in $m-1$ by using the semiclassical conformal block. In the gravity side, we compute the entanglement wedge cross section in the backreacted geometry by cosmic branes with tension $T_m$ which are anchored at the AdS boundary.  Comparing both results we find a perfect agreement, showing the duality works with the first order correction in $m-1$.

}

\begin{document}

\maketitle

%%%%%%%%%%%%%%%%%%%%%%%%%%%%%%%%%%%%%%
%    INTRODUCTION
%%%%%%%%%%%%%%%%%%%%%%%%%%%%%%%%%%%%%%
\section{Introduction}
The AdS/CFT correspondence (or gauge/gravity duality)~\cite{Maldacena:1997re, Gubser:1998bc, Witten:1998qj} is an interesting duality between gravity theories and conformal field theories (CFTs). It provides a new viewpoint to better understand field theories in terms of geometric quantities. In recent years, a remarkable perspective on this duality has been developed in the quantum entanglement and information theory. 

Considering the quantification of entanglement is the necessary foundation in quantum information theories, which basically corresponds to studying \textit{entanglement measure}. 
One important and well-studied entanglement measure in the gauge/gravity duality framework is the entanglement entropy. The Ryu-Takayanagi formula \cite{Ryu:2006bv, Ryu:2006ef} gives us a hint of the emergence of spacetime from the entanglement entropy in the dual conformal field theories (e.g., see~\cite{Nishioka:2009un, VanRaamsdonk:2010pw, Nozaki:2012zj, Lin:2014hva, Hayden:2016cfa}).

The entanglement entropy is suitable for measuring the quantum entanglement of \textit{pure states}, however, it is not a good measure for \textit{mixed states} in that the entanglement entropy could be nonzero even though the two subsystems are not entangled (e.g., see~\cite{Horodecki:2009zz}). Hence, it is important to construct other entanglement measure quantities in order to investigate mixed states. From the holographic point of view, new geometric objects describing mixed states are now required, which are expected to be different from usual minimal surfaces in the Ryu-Takayanagi formula. The entanglement wedge cross section \cite{Takayanagi:2017knl, Nguyen:2017yqw} is suggested as such an object in holography, which is defined by minimal surfaces in the entanglement wedge. The entanglement wedge \cite{Czech:2012bh, Wall:2012uf, Headrick:2014cta} is a bounded region of the bulk spacetime dual to a reduced density matrix. Since the reduced density matrix is  a mixed state in general, the entanglement wedge cross section is expected to be the holographic dual of some entanglement measures for the mixed states. The more detailed description of the entanglement wedge cross section will be reviewed in section \ref{sec3}. There are various proposals of entanglement measures for mixed states in the CFTs as the dual of entanglement wedge cross section: entanglement of purification  \cite{Takayanagi:2017knl, Nguyen:2017yqw}, logarithmic negativity \cite{Kudler-Flam:2018qjo}, odd entanglement entropy \cite{Tamaoka:2018ned}, and reflected entropy \cite{Dutta:2019gen}.  See also recent studies of the entanglement wedge cross section in \cite{Bao:2017nhh, Hirai:2018jwy, Espindola:2018ozt, Bao:2018gck, Umemoto:2018jpc, Yang:2018gfq, Bao:2018fso, Agon:2018lwq, Bao:2018pvs,  Caputa:2018xuf, Liu:2019qje, Kudler-Flam:2019oru, BabaeiVelni:2019pkw, Du:2019emy, Jokela:2019ebz, Guo:2019pfl, Bao:2019wcf, Harper:2019lff, Kudler-Flam:2019wtv,  Kusuki:2019rbk, Kusuki:2019zsp, Wang:2019ued, Umemoto:2019jlz, Suzuki:2019xdq}.

Hereafter, we focus on the reflected entropy $S_R(A: B)$, which is the entanglement entropy of a canonically purified state generated from a given density matrix $\rho_{AB}$ on a bipartite Hilbert space $\mathcal{H}_A\otimes\mathcal{H}_B$. Motivated by the duality between the thermofield double state and the eternal AdS black hole \cite{Maldacena:2001kr}, the authors of \cite{Dutta:2019gen} proposed the following duality
\begin{align}
S_R(A: B)=2E_W(A: B)+\mathcal{O}(G_N^0),\label{d1}
\end{align}
where $E_W(A: B)$ is the entanglement wedge cross section, {which is the area of minimal cross section in the entanglement wedge divided by $4G_N$,} and $G_N$ is the gravitational constant. With the reduced density matrix $\rho_{AB}$ of the ground state in the 2d holographic CFTs on two disjoint intervals $A$ and $B$, the duality (\ref{d1}) was explicitly checked \cite{Dutta:2019gen}. The duality with the time evolution by a quench was also studied in \cite{Kusuki:2019rbk, Wang:2019ued}.

Based on the replica trick in the bulk \cite{Lewkowycz:2013nqa, Faulkner:2018faa}, the duality (\ref{d1}) was established as \cite{Dutta:2019gen}
\begin{align}
\lim_{n\to1}S_n(AA^\star)_{\psi_m}=2E_{mW}(A: B)+\mathcal{O}(G_N^0),\label{d2}
\end{align}
by assuming the replica and time reflection symmetry in the bulk and the GKP-Witten relation \cite{Gubser:1998bc, Witten:1998qj}. Here, $\lim_{n\to1}S_n(AA^\star)_{\psi_m}$ is the reflected entropy for $\rho_{AB}^m$,  and $E_{mW}(A: B)$ is the entanglement wedge cross section in the quotient spacetime $\mathcal{B}_m/\mathbb{Z}_m$, which is used to compute the holographic R\'{e}nyi entropy \cite{Lewkowycz:2013nqa, Faulkner:2013yia}. We will explain the detail of $S_n(AA^\star)_{\psi_m}$ and $E_{mW}(A: B)$ in the main context. 

{Our motivations to consider $m \ne 1$ are three folds. The first motivation comes from the holographic duality.  From the gravity side, $m \ne 1$ has a clear meaning that we need to consider the back-reaction of the cosmic brane~\cite{Dong:2016fnf} when we compute the entanglement wedge cross section. According to the holographic duality, there must be a dual quantity in field theory side, which was proposed to be the reflected entropy of $\rho_{AB}^m$~\cite{Dutta:2019gen}. Even though the field theory meaning of this quantity is not clear for now,\footnote{{The holographic duality says that there exist dual quantities in gravity and field theory side, but a simple quantity in one side may not be necessarily a simple quantity in the other side.}} it is a well defined and important question to ask if the aforementioned holographic proposal (Eq.~\eqref{d2} for $m\ne1$) is valid. In this work, we consider $m \sim 1$ for technical reasons so only an $(m-1)$ correction. 
Second, from the replica-trick perspective, we first formulate $S_n(AA^\star)_{\psi_m}$ for  $n\in\mathbb{Z}^+$ and $m\in2\mathbb{Z}^+$ as explained in section 2. Then, we consider an analytic continuation with $n\to1$ and $m\to1$ to compute $S_R(A: B)$. It means $S_n(AA^\star)_{\psi_m}$ needs to be defined well for all $m$ including for small ($m-1$).
Third, another motivation to consider the generalization by $m$ in the field theory side is related to eigenvalues of $\rho_{AB}$. 
Computing  $\lim_{m\to1}S_n(AA^\star)_{\psi_m}$ with all $n$ will provide us eigenvalues of $\rho_{AA^\star}$ because $\lim_{m\to1}S_n(AA^\star)_{\psi_m}$ is the $n$-th R\'{e}nyi entropy of the reduced density matrix $\rho_{AA^\star}$.
Furthermore, one may investigate the eigenvalues of $\rho_{AB}$ from $\lim_{m\to1}S_n(AA^\star)_{\psi_m}$ or the eigenvalues of $\rho_{AA^\star}$ via the construction of $\rho_{AA^\star}$ from $\rho_{AB}$. However, it is not certain that we can determine eigenvalues of $\rho_{AB}$ uniquely from $\lim_{m\to1}S_n(AA^\star)_{\psi_m}$ only because there is a possibility that the inverse map from $\rho_{AA^\star}$ to $\rho_{AB}$ is not uniquely determined.
Thus, we expect that $S_n(AA^\star)_{\psi_m}$ has more information about the eigenvalues of $\rho_{AB}$ than $\lim_{m\to1}S_n(AA^\star)_{\psi_m}$, and this is one motivation to consider the generalization by $m$.}

Note that the quotient spacetime $\mathcal{B}_m/\mathbb{Z}_m$ has conical singularities, which are fixed points of the $\mathbb{Z}_m$ symmetry in $\mathcal{B}_m$, and these singularities can be interpreted as cosmic branes with tension $T_m = \frac{m-1}{4mG_N}$ \cite{Dong:2016fnf}. As with the holographic R\'{e}nyi entropy, these cosmic branes produce the $m$-dependence of  $E_{mW}(A: B)$ by backreaction to the bulk geometry \cite{Dutta:2019gen, Wang:2019ued}. The geometry with the backreaction from a single cosmic brane homologous to a disk was studied in \cite{Hung:2011nu}. A construction procedure of the bulk geometry with the backreaction for two intervals was developed in \cite{Faulkner:2013yia}. Especially, the author of \cite{Dong:2016fnf} computed the area of a single cosmic brane with the backreaction from the other cosmic brane at  first order in $m-1$, giving non-vanishing tension of cosmic branes.  One can also introduce the backreaction by considering $n\ne1$. In particular, the R\'{e}nyi reflected entropy with $n=1/2$ and its bulk dual with the backreaction  were studied in \cite{Kusuki:2019zsp} for the holographic dual of logarithmic negativity. 

{For general value of $m$ and the configuration of the subsystems $A$ and $B$, although the holographic duality (\ref{d2}) was established by the Lewkowycz-Maldacena type derivation, an explicit computation of (\ref{d2}) is not simple because a construction of $\mathcal{B}_m/\mathbb{Z}_m$ is complicated. Moreover, the Lewkowycz-Maldacena type derivation is based on the GKP-Witten relation, but the proof of the GKP-Witten relation is generally very difficult. Hence, an explicit calculation of (\ref{d2}), at least by a simple example, is important for the consistency check of the duality.}

In this work, we explicitly compute and show (\ref{d2}) with the two disjoint intervals $A$ and $B$ at first order in $m-1$. We evaluate $\lim_{n\to1}S_n(AA^\star)_{\psi_m}$ for the reduced density matrix  of the ground state in the 2d holographic CFTs as well as $S_R(A: B)$ studied  in \cite{Dutta:2019gen}. The entanglement wedge cross section $E_{mW}(A: B)$ with the small backreaction can be obtained by a method in  \cite{Dong:2016fnf} for the holographic R\'{e}nyi entropy. By comparing the two results, we find an exact agreement, which means an explicit check of (\ref{d2}) up to first order in $m-1$.

%We apply this computation method to evaluate $E_{mW}(A: B)$ at first order in $m-1$ and show that the duality \eqref{d2} is also valid with the first order correction.
%In this paper showing an explicit check of \eqref{d2} by a simple and computable example with the first order correction, we lay the foundation stone of more complicated or general cases.

The organization of this paper is as follows.
In section \ref{sec2}, we review the reflected entropy and compute the left hand side of (\ref{d2}) at first order in $m-1$.  In section \ref{sec3}, we derive the right hand side of (\ref{d2}) with the small backreaction and check the duality (\ref{d2}). We conclude in section \ref{sec4}.

%%%%%%%%%%%%%%%%%%%%%%%%%%%%%%%%%%%%%%%%%
%
%        Section: Reflected entropy for $\rho_{AB}^m$ at first order in $m-1$
%
%%%%%%%%%%%%%%%%%%%%%%%%%%%%%%%%%%%%%%%%%

\section{Reflected entropy with the first order correction}\label{sec2}
In this section we study the reflected entropy for $\rho_{AB}^{m}$ with two disjoint intervals $A$ and $B$ in the 2d holographic CFTs, where $\rho_{AB}$ is the reduced density matrix of the ground state.
For this purpose, we first review the reflected entropy for finite dimensional Hilbert spaces and generalize it for continuous field theories using the replica trick~\cite{Dutta:2019gen}. In particular, as a functional calculation tool, we will express the reflected entropy in terms of the twist operators and compute it up to first order correction in the replica index $m$ using a perturbative expansion of the semiclassical conformal block.

\subsection{Some formalism}
First of all, we review the reflected entropy based on \cite{Dutta:2019gen, Kusuki:2019rbk, Kusuki:2019zsp} for finite dimensional Hilbert spaces. Consider a positive-semidefinite density matrix $\rho_{AB}$ on a Hilbert space $\mathcal{H}_A\otimes\mathcal{H}_B$:
\begin{align}
\rho_{AB}:=\sum_{a}p_a|\phi_a\rangle\langle \phi_a|_,\label{DM}
\end{align}
where $|\phi_a\rangle$ is an orthogonal and normalized basis of $\mathcal{H}_A\otimes\mathcal{H}_B$, and $p_a$ are nonnegative eigenvalues. We normalize (\ref{DM}) as $\text{Tr}\rho_{AB}=\sum_{a}p_a=1$. By choosing appropriate bases $|i_a\rangle_A$ of $\mathcal{H}_A$ and $|i_a\rangle_B$ of $\mathcal{H}_B$, we can construct a Schmidt decomposition of $|\phi_a\rangle$ (see, for example, \cite{Horodecki:2009zz}):
\begin{align}
|\phi_a\rangle=\sum_i\sqrt{l^i_a}|i_a\rangle_A|i_a\rangle_B,\label{psin}
\end{align}
where $l^i_a$ is a nonnegative value with the normalization $\sum_il^i_a=1$. Substituting (\ref{psin}) into (\ref{DM}), we obtain
\begin{align}
\rho_{AB}=\sum_{a, i, j}p_a\sqrt{l^i_al^j_a}|i_a\rangle_A|i_a\rangle_B\langle j_a|_A\langle j_a|_B.
\end{align}

Interpreting $\langle j_a|_A$ and $\langle j_a|_B$ as states {$|j_a\rangle_{A^\star}$ and $|j_a\rangle_{B^\star}$} on Hilbert spaces $\mathcal{H}^\star_A$ and $\mathcal{H}^\star_B$ respectively, we can define a state $|\sqrt{\rho_{AB}}\rangle$ on $\mathcal{H}_A\otimes\mathcal{H}_B\otimes\mathcal{H}^\star_A\otimes\mathcal{H}^\star_B$ as
{\begin{align}
|\sqrt{\rho_{AB}}\rangle := \sum_{a, i, j}\sqrt{p_al^i_al^j_a}|i_a\rangle_A|i_a\rangle_B|j_a\rangle_{A^\star}|j_a\rangle_{B^\star}.\label{SDM}
\end{align}}
One can easily show that $|\sqrt{\rho_{AB}}\rangle$ represents a purification of $\rho_{AB}$ as follows
\begin{align}
\text{Tr}_{\mathcal{H}^\star_A\otimes\mathcal{H}_B^\star}|\sqrt{\rho_{AB}}\rangle\langle\sqrt{\rho_{AB}}|=\rho_{AB} \,.
\end{align}
Then, with the state \eqref{SDM}, the reflected entropy $S_R(A: B)$ for $\rho_{AB}$ is defined by
\begin{align}
S_R(A: B) := \,&-\text{Tr}_{\mathcal{H}_A\otimes\mathcal{H}^\star_A}\left[\rho_{AA^\star}\log\rho_{AA^\star}\right],\notag\\
\rho_{AA^\star} := \,&\text{Tr}_{\mathcal{H}_B\otimes\mathcal{H}^\star_B}|\sqrt{\rho_{AB}}\rangle\langle\sqrt{\rho_{AB}}|.\label{RE}
\end{align}
Note that the reflected entropy in \eqref{RE} follows the form of the Von Neumann entropy. In other words, we can understand the reflected entropy $S_R(A: B)$ as the entanglement entropy of the reduced density matrix $\rho_{AA^\star}$.

\subsection{Replica trick for the reflected entropy}
In this section, we rewrite the definition of $S_R(A: B)$ for continuous field theories by the replica trick.
After giving the expression of reflected entropy in terms of partition functions, we will reformulate it with the twist operators.
To formulate $S_R(A: B)$ for continuous field theories by the replica trick, $S_R(A: B)$ is generalized by two replica indices  $n$ and $m$~\cite{Dutta:2019gen}. In terms of the replica index, $\rho_{AB}$ in (\ref{DM}) is generalized by $m$ as
\begin{align}
\begin{split}
\rho_{AB} ^{m} :=&\sum_{a}p_a^m|\phi_a\rangle\langle \phi_a|_,\label{DDM} 
= \sum_{a, i, j}p_a^{m}\sqrt{l^i_al^j_a}|i_a\rangle_A|i_a\rangle_B\langle j_a|_A\langle j_a|_B.
\end{split}
\end{align}
where, \eqref{psin} is used in the last equality. Accordingly, $|\sqrt{\rho_{AB}}\rangle$ in (\ref{SDM}) is generalized as
{\begin{align}
|\rho_{AB}^{m/2}\rangle:=&\sum_{a, i, j}p_a^{m/2}\sqrt{l^i_al^j_a} \,\, |i_a\rangle_A|i_a\rangle_B|j_a\rangle_{A^\star}|j_a\rangle_{B^\star},  \notag\\
|\psi_m\rangle:=&\frac{1}{\sqrt{\text{Tr}\rho_{AB}^m}}|\rho_{AB}^{m/2}\rangle, \label{psim}
\end{align}}
where $|\psi_m\rangle$ is a purification of $\rho_{AB}^m$ in \eqref{DDM} with the normalization:
\begin{align}
\text{Tr}_{\mathcal{H}^\star_A\otimes\mathcal{H}_B^\star}|\psi_m\rangle\langle\psi_m| = \frac{\rho_{AB}^m}{\text{Tr}\rho_{AB}^m} \,.
\end{align}
Then, finally the reflected entropy $S_R(A: B)$ (\ref{RE}) is generalized by $n$ and $|\psi_m\rangle$ \eqref{psim} as
\begin{align}
S_n(AA^\star)_{\psi_m}:=\,\,&\frac{1}{1-n}\log\text{Tr}_{\mathcal{H}_A\otimes\mathcal{H}^\star_A}\left(\rho_{AA^\star}^{(m)}\right)^n,\notag\\
\rho_{AA^\star}^{(m)}:=\,\,&\text{Tr}_{\mathcal{H}_B\otimes\mathcal{H}^\star_B}|\psi_m\rangle\langle\psi_m|, \label{SNEQ}
\end{align}
where $S_n(AA^\star)_{\psi_m}$ is the $n^\text{th}$ R\'{e}nyi entropy of the reduced density matrix $\rho_{AA^\star}^{(m)}$. When $n\to1$ and $m\to1$, $S_n(AA^\star)_{\psi_m}$ reduces to $S_R(A: B)$
\begin{align}
\lim_{n, m\to1}S_n(AA^\star)_{\psi_m}=S_R(A: B).
\end{align}

Introducing partition functions $Z_{n, m}$ as
\begin{align}
Z_{n, m}:=\text{Tr}_{\mathcal{H}_A\otimes\mathcal{H}^\star_A}\left(\text{Tr}_{\mathcal{H}_B\otimes\mathcal{H}^\star_B}\left|\rho_{AB}^{m/2}\right\rangle\left\langle\rho_{AB}^{m/2}\right|\right)^n,\label{PF1}
\end{align}
$S_n(AA^\star)_{\psi_m}$ in \eqref{SNEQ} can be expressed by
\begin{align}
S_n(AA^\star)_{\psi_m}=\frac{1}{1-n}\log\frac{Z_{n, m}}{(Z_{1, m})^n}. \label{PF2}
\end{align}
The authors of \cite{Dutta:2019gen} gave a prescription for $S_n(AA^\star)_{\psi_m}$ in CFTs. In particular, they formulated $Z_{n, m}$ by a path integral on a replica manifold for $n\in\mathbb{Z}^+$ and $m\in2\mathbb{Z}^+$. {The condition $m\in2\mathbb{Z}^+$ is related to the replica manifold of $|\rho_{AB}^{m/2}\rangle$. The number of replica sheets in the replica manifold of $|\rho_{AB}^{m/2}\rangle$ is $m/2$, and thus, $m/2$ must be a positive integer.} By using an analytic continuation of $m$ and $n$, they evaluated the reflected entropy by $\lim_{n, m\to1}S_n(AA^\star)_{\psi_m}$ in the 2d holographic CFTs. 

\paragraph{Constructing the replica manifold for  $Z_{n, m}$:} We review how to construct the replica manifold for $Z_{n, m}$ (\ref{PF1}) with the reduced density matrix $\rho_{AB}$ of the ground state in 2d CFTs. Here for a vivid example of it, we give $n=2, \, m=4$ case.  
Let us start with the manifold of $|\rho_{AB}^{2}\rangle$ composing a basic building block of $Z_{2, 4}$: $|\rho_{AB}^{2}\rangle\langle\rho_{AB}^{2}|$. The overall structure of the manifold of $|\rho_{AB}^{2}\rangle$ is the same as that of the density matrix $\rho_{AB}^{2}$. This is due to the resemblance between $\rho_{AB}^{m/2}$ in \eqref{DDM} and $|\rho_{AB}^{m/2}\rangle$ in \eqref{psim}. Only difference between them is the Hilbert spaces in which two intervals live in, following explanation near \eqref{DM} and \eqref{SDM}, %$|\rho_{AB}^{m/2}\rangle$ on $\mathcal{H}_A\otimes\mathcal{H}_B\otimes\mathcal{H}^\star_A\otimes\mathcal{H}^\star_B$ can be interpreted as the density matrix $\rho_{AB}^{m/2}$ on $\mathcal{H}_A\otimes\mathcal{H}_B$, namely
the density matrix $\rho_{AB}^{m/2}$ on $\mathcal{H}_A\otimes\mathcal{H}_B$ can be interpreted as the pure state $|\rho_{AB}^{m/2}\rangle$ on $\mathcal{H}_A\otimes\mathcal{H}_B\otimes\mathcal{H}^\star_A\otimes\mathcal{H}^\star_B$, namely
\begin{align}
\begin{split}
|\rho_{AB}^{2}\rangle   \in \, \mathcal{H}_A\otimes\mathcal{H}_B\otimes\mathcal{H}^\star_A\otimes\mathcal{H}^\star_B \,. \label{DIF}
\end{split}
\end{align}
\begin{figure}[]
\centering
     {\includegraphics[width=15cm]{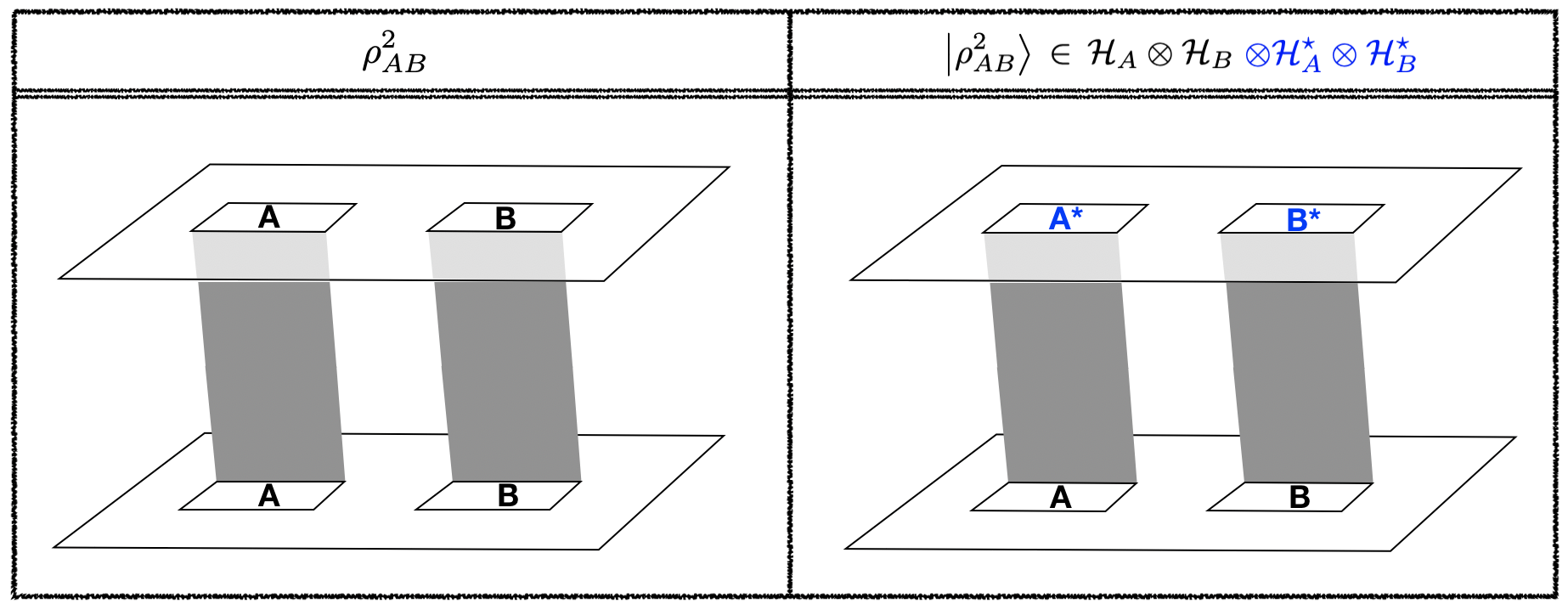} \label{}}
 \caption{Replica manifold of $\rho_{AB}^{2}$ and $|\rho_{AB}^{2}\rangle$. The difference of their Hilbert spaces \eqref{DIF} is marked in blue color.}\label{MANIFOLDFIG1}
\end{figure} 
Explicit shape of their manifold is displayed in Fig. \ref{MANIFOLDFIG1}.

\begin{figure}[]
\centering
     {\includegraphics[width=15cm]{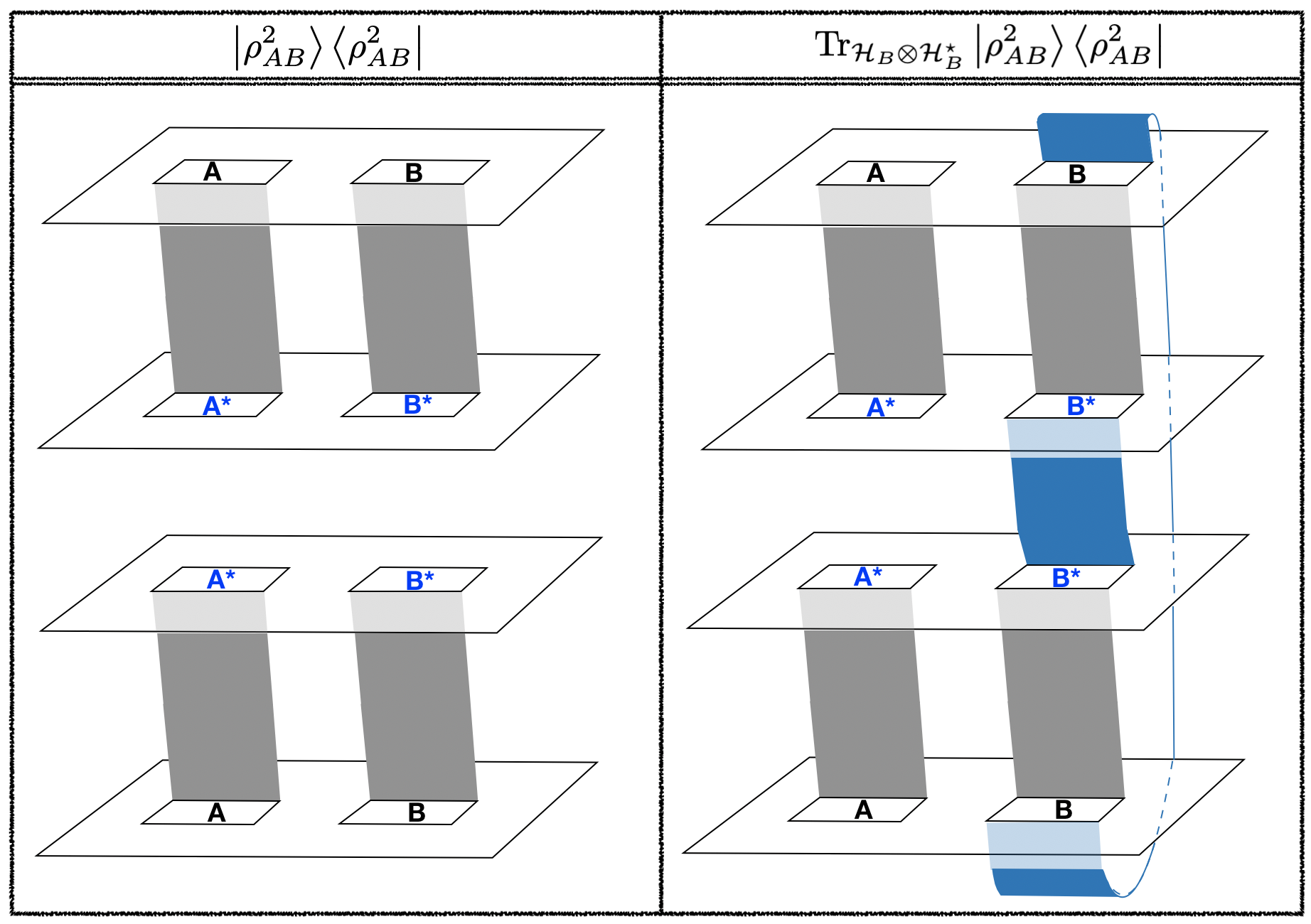} \label{}}
 \caption{Replica manifold of $|\rho_{AB}^{2}\rangle\langle\rho_{AB}^{2}|$ and $\text{Tr}_{\mathcal{H}_B\otimes\mathcal{H}^\star_B}|\rho_{AB}^{2}\rangle\langle\rho_{AB}^{2}|$. Note that the hermitian conjugate $\langle\rho_{AB}^{2}|$ has a switched the position on ($\mathcal{H}^\star_A, \, \mathcal{H}^\star_B$) in comparison with $|\rho_{AB}^{2}\rangle$ in Fig. \ref{MANIFOLDFIG1}. The trace $\text{Tr}_{\mathcal{H}_B\otimes\mathcal{H}^\star_B}$ is done by gluing intervals $B$ (and $B^\star$) on different sheets.}\label{MANIFOLDFIG2}
\end{figure}

Using the description of $|\rho_{AB}^{2}\rangle$ above, we can make the replica manifold of $|\rho_{AB}^{2}\rangle\langle\rho_{AB}^{2}|$ and the trace of it, $\text{Tr}_{\mathcal{H}_B\otimes\mathcal{H}^\star_B}|\rho_{AB}^{2}\rangle\langle\rho_{AB}^{2}|$, as shown in Fig. \ref{MANIFOLDFIG2}.
Note that the positions of ($\mathcal{H}_A, \, \mathcal{H}_B$) and ($\mathcal{H}^\star_A, \, \mathcal{H}^\star_B$) in the replica manifold of the hermitian conjugate $\langle\rho_{AB}^{2}|$ are switched in comparison with $|\rho_{AB}^{2}\rangle$\footnote{The trace $\text{Tr}_{\mathcal{H}_B\otimes\mathcal{H}^\star_B}$ is done by gluing intervals $B$ (and $B^\star$) on different sheets.}.

\begin{figure}[]
\centering
     {\includegraphics[width=15cm]{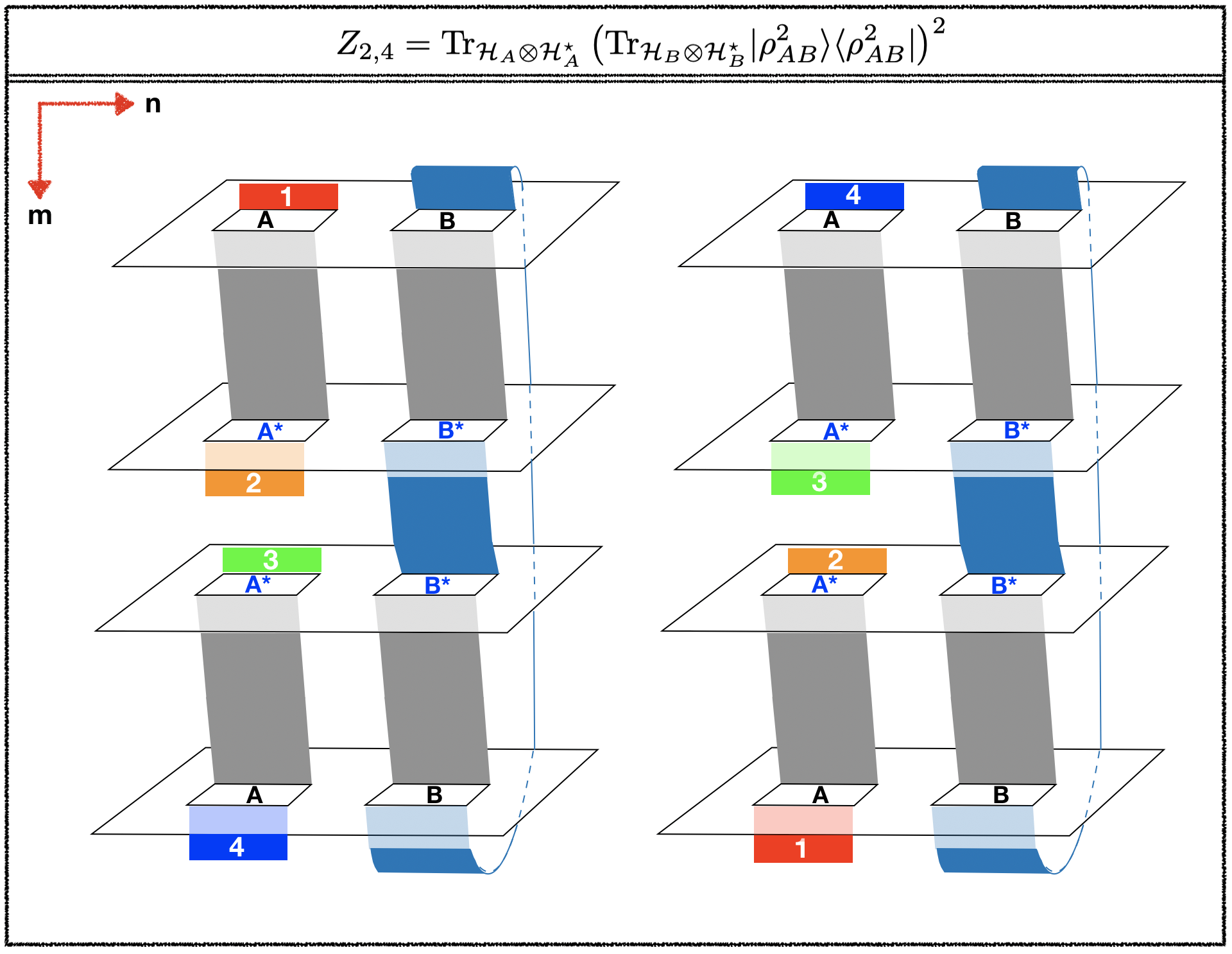} \label{}}
 \caption{Replica manifolds of $Z_{n, m}=\text{Tr}_{\mathcal{H}_A\otimes\mathcal{H}^\star_A}\left(\text{Tr}_{\mathcal{H}_B\otimes\mathcal{H}^\star_B}|\rho_{AB}^{m/2}\rangle\langle\rho_{AB}^{m/2}|\right)^n$ with replica indices $n=2$ and $m=4$. The small colored panel on each sheet have a numbering mark on them. It represents the connections between the same numbered panel (or the same colored panel).}\label{MANIFOLDFIG3}
\end{figure}
One can notice that the small colored panel on each sheet in Fig. \ref{MANIFOLDFIG3} have a numbering mark on them. It represents the connections between the same numbered panel (or the same colored panel). The way of gluing them is determined when we introduce $n$ as follows. For instance, we have two $\text{Tr}_{\mathcal{H}_B\otimes\mathcal{H}^\star_B}|\rho_{AB}^{2}\rangle\langle\rho_{AB}^{2}|$ before taking the trace $\text{Tr}_{\mathcal{H}_A\otimes\mathcal{H}^\star_A}$ in $Z_{2, 4}$. This means that the inner product is evaluated between the bra state $\langle\rho_{AB}^{2}|$ from one piece of $\left(\text{Tr}_{\mathcal{H}_B\otimes\mathcal{H}^\star_B}|\rho_{AB}^{2}\rangle\langle\rho_{AB}^{2}|\right)^2$ and the ket state $|\rho_{AB}^{2}\rangle$ from another. This procedure correspond to how the red (or orange) colored panel get glued together in Fig. \ref{MANIFOLDFIG3}. After doing this procedure, the remaining trace operation ($\text{Tr}_{\mathcal{H}_A\otimes\mathcal{H}^\star_A}$) acts on $\left(\text{Tr}_{\mathcal{H}_B\otimes\mathcal{H}^\star_B}|\rho_{AB}^{2}\rangle\langle\rho_{AB}^{2}|\right)^2$. In terms of the replica manifold desctiption, it can be viewed as a connecting green (and blue) colored panel in Fig. \ref{MANIFOLDFIG3}.

\paragraph{Twist operator representation of $S_n(AA^\star)_{\psi_m}$:}
In the same way as the entanglement entropy in 2d CFTs \cite{Calabrese:2004eu, Calabrese:2009qy}, the path integral representation of $S_n(AA^\star)_{\psi_m}$ on the replica manifold  can be expressed by correlation functions of the twist operators \cite{Dutta:2019gen}:
\begin{align}
S_n(AA^\star)_{\psi_m}=\frac{1}{1-n}\log\frac{\left\langle\sigma_{g_A}(x_1)\sigma_{g_A^{-1}}(x_2)\sigma_{g_B}(x_3)\sigma_{g_B^{-1}}(x_4)\right\rangle_{CFT^{\otimes mn}}}{\left(\left\langle\sigma_{g_m}(x_1)\sigma_{g_m^{-1}}(x_2)\sigma_{g_m}(x_3)\sigma_{g_m^{-1}}(x_4)\right\rangle_{CFT^{\otimes m}}\right)^n},\label{RRE}
\end{align}
where we take the two intervals $A=[x_1, x_2]$ and $B=[x_3, x_4]$ with $x_1<x_2<x_3<x_4$, and  $CFT^{\otimes mn}$ is the product theory on 2d flat spacetime, which contains $mn$ replica fields for the $mn$ replica sheets as in Fig. \ref{MANIFOLDFIG3}. The twist operators $\sigma_{g_A}$, $\sigma_{g_A^{-1}}$, $\sigma_{g_B}$, and $\sigma_{g_B^{-1}}$ are defined such that the replica fields satisfy boundary conditions around the twist operators, and these boundary conditions  are determined by the connection between the replica sheets. However, unlike above twist operators, the twist operators  $\sigma_{g_m}$ and $\sigma_{g_m^{-1}}$ are defined to be the cyclic connections between $m$ replica sheets, namely, ($\sigma_{g_m}, \sigma_{g_m^{-1}}$) can only be applied to $m$-direction. Thus, when $n=1$, these three operators($\sigma_{g_A}, \sigma_{g_B}, \sigma_{g_m}$) are equal,
\begin{align}
\sigma_{g_A}=\sigma_{g_B}=\sigma_{g_m} \;\;(n=1). \label{EQUAL}
\end{align}

Note that the product theory $CFT^{\otimes mn}$ in (\ref{RRE}) is not an orbifold theory\footnote{As explained in \cite{Dutta:2019gen}, the twist operators in (\ref{RRE}) without orbifolding are not quite local operators, however, we can define the OPE (\ref{OPE}). See \cite{Dutta:2019gen, Balakrishnan:2017bjg} for more details.}. Thus, $\sigma_{g_A}$ and $\sigma_{g_B}$ are not identified at $n\ne1$, and the OPE between $\sigma_{g_A^{-1}}$ and $\sigma_{g_B}$ includes not the unit operator but rather a twist operator $\sigma_{g_Bg_A^{-1}}$,
\begin{align}
\sigma_{g_A^{-1}}\sigma_{g_B} \,\to\, \sigma_{g_Bg_A^{-1}} \,+\, \cdots.\label{OPE}
\end{align} 

The conformal dimensions $h_{g_A^{-1}}$ of $\sigma_{g_A^{-1}}$ and $h_{g_B}$ of $\sigma_{g_B}$ are \cite{Dutta:2019gen}\footnote{For the twist operators, $h_{g_A^{-1}} = h_{g_A}$ and $h_{g_B^{-1}} = h_{g_B}$.}
\begin{align}
h_{g_A^{-1}}=h_{g_B}=\frac{nc}{24}\left(m-\frac{1}{m}\right). \label{ht}
\end{align}
These values can be explained as follows. The replica manifold in Fig. \ref{MANIFOLDFIG3} includes $n$ cyclic loops which connect the $m$ replica sheets through $B$ and $B^{\star}$. Hence, we may say that the conformal dimension of $\sigma_{g_B}$ is $h_{g_B} = n \, h_m$, where $h_m:=\frac{c}{24}(m-1/m)$ is the conformal dimension of usual twist operators for $m$ replica sheets \cite{Calabrese:2004eu, Calabrese:2009qy}. The same is true for $h_{g_A^{-1}}$. 
On the other hand, the conformal dimension $h_{g_Bg_A^{-1}}$ of $\sigma_{g_Bg_A^{-1}}$ is given as~\cite{Dutta:2019gen}
\begin{align}
h_{g_Bg_A^{-1}}=\frac{2c}{24}\left(n-\frac{1}{n}\right).\label{ht2}
\end{align}
\begin{figure}[]
\centering
     {\includegraphics[width=15cm]{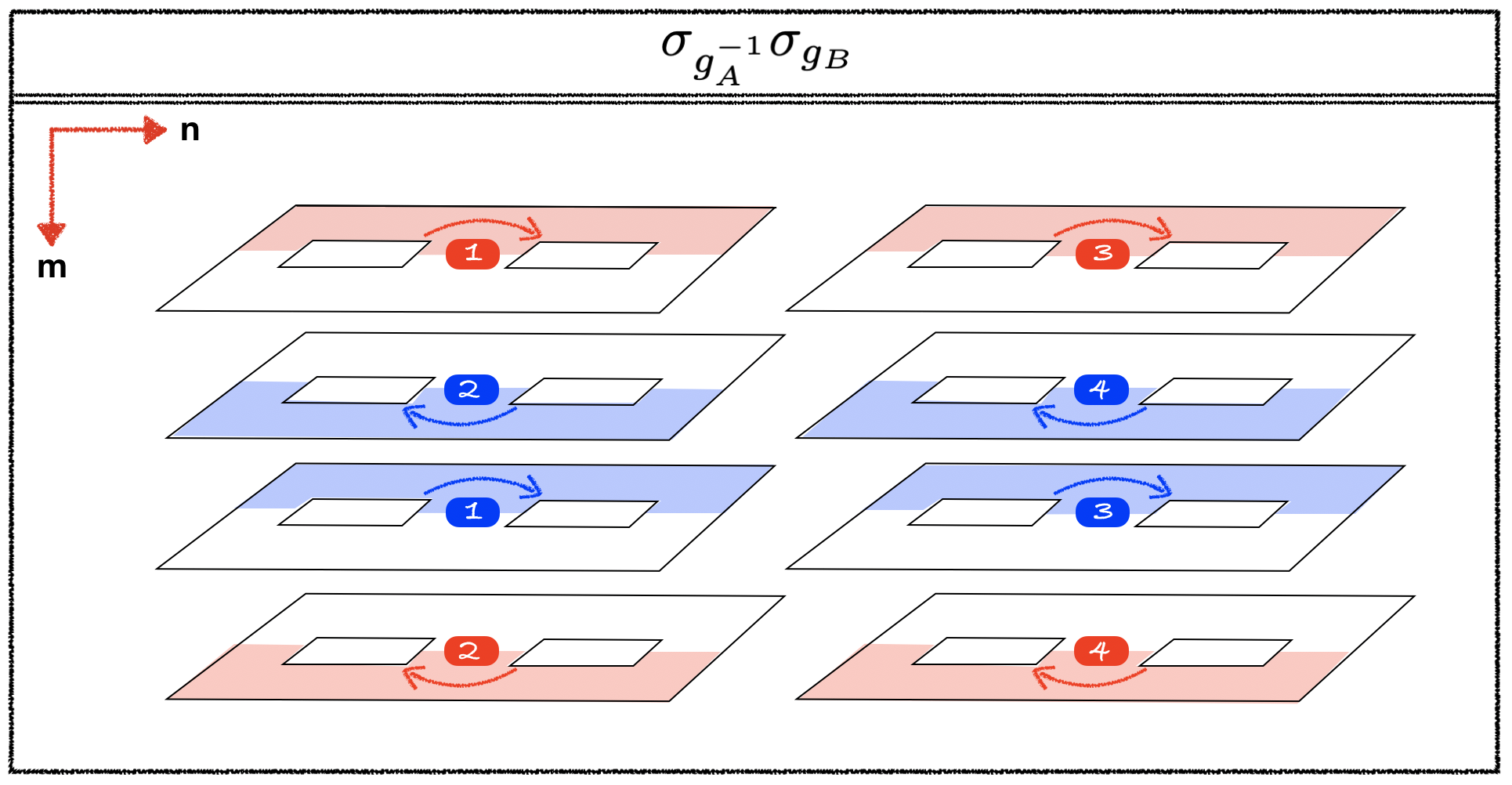}}
 \caption{$\sigma_{g_A^{-1}}\sigma_{g_B} \sim \sigma_{g_Bg_A^{-1}}$ in the replica manifold $Z_{2, 4}$. Two cyclic loops are represented in red and blue colored arrows with numbering. The numbering follows the sequence of connection between each replica sheet in Fig. \ref{MANIFOLDFIG3}.} \label{MANIFOLDFIG4}
\end{figure}This conformal dimension \eqref{ht2} is a consequence of two things: i) the way of satisfying boundary conditions of twist operator, related to the rotation around the end points of intervals~\cite{Calabrese:2004eu,Calabrese:2009qy}, ii) specific intertwined structure of replica manifold in Fig. \ref{MANIFOLDFIG3}. Here, we give an example of $\sigma_{g_A^{-1}}\sigma_{g_B}$ with $n=2 \,, m=4$ to explain how \eqref{ht2} can be obtained. In terms of twist operators, boundary conditions in the replica manifold are satisfied by performing a rotation around end points of intervals: for instance, anti-twist operator $\sigma_{g_A^{-1}}$ is acting on the right point of interval $A$ and a twist operator $\sigma_{g_B}$ does on the left point of interval $B$. By combining those two twist operator's rotational effect with the intertwined structure of replica manifold in Fig. \ref{MANIFOLDFIG3}, we display how  $\sigma_{g_A^{-1}}\sigma_{g_B}$ relates the sheets in the manifolds in Fig. \ref{MANIFOLDFIG4}.
Note that the rotation for $\sigma_{g_A^{-1}}\sigma_{g_B}$ is only on the half region of each sheet, which is represented as shaded regions in red (or blue) color. Then, we can recognize that there are two cyclic loops in Fig. \ref{MANIFOLDFIG4}. One loop is represented with red arrows with numbering, and the other loop is with blue arrows\footnote{One can easily check this numbering with Fig. \ref{MANIFOLDFIG3}.}. 
Since four half pieces of sheets correspond to two complete sheets, each loop can be regarded as the usual rotations in $n=2$ manifold as in the Renyi entropy. Thus, the conformal dimension of $\sigma_{g_Bg_A^{-1}}$ is $h_{g_Bg_A^{-1}} = 2 \, h_n$ where $h_n$ is given in \eqref{cnm}\footnote{See also explanation by group elements $g_B$ and $g_A^{-1}$ in \cite{Dutta:2019gen}.}.

\subsection{Reflected entropy in the 2d holographic CFTs up to first order in $m-1$}
Using conformal dimensions of twist operators \eqref{ht} and \eqref{ht2},  we compute correlation functions in \eqref{RRE}. 
In any 2d CFTs with the Virasoro symmetry, the four point function $\left\langle\sigma_{g_A}(x_1)\sigma_{g_A^{-1}}(x_2)\sigma_{g_B}(x_3)\sigma_{g_B^{-1}}(x_4)\right\rangle_{CFT^{\otimes mn}}$ can be expanded by conformal blocks in $t$-channel  (see, for example, \cite{Ginsparg:1988ui, Hartman:2013mia,  Perlmutter:2015iya, Tamaoka:2018ned})
\begin{align}
&\left\langle\sigma_{g_A}(x_1)\sigma_{g_A^{-1}}(x_2)\sigma_{g_B}(x_3)\sigma_{g_B^{-1}}(x_4)\right\rangle_{CFT^{\otimes mn}}\notag\\
=& \,\, \frac{1}{(x_4-x_1)^{2(h+\bar{h})}(x_3-x_2)^{2(h+\bar{h})}} \,\, \sum_{p}C_{ABp}^2 \,\,\, \mathcal{F}(mnc, h, h_p, 1-z) \,\, \mathcal{F}(mnc, \bar{h}, \bar{h}_p, 1-\bar{z}),\label{cbe}
\end{align}
where $h=\bar{h}=\frac{nc}{24}\left(m-\frac{1}{m}\right)$ are the conformal dimensions of the twist operators (\ref{ht}), the sum $\sum_p$ is over primary operators $\mathcal{O}_p$ with the conformal dimensions $h_p$ and $\bar{h}_p$\footnote{In this paper, we mainly consider the exchange of the twist operators with $h_p=\bar{h}_p$.}, and $C_{ABp}$ is the OPE coefficient of three point functions. In addition, $\mathcal{F}$ is the Virasoro conformal block and $mnc$ represents the central charge of $CFT^{\otimes mn}$. In our set up of the two intervals, the cross ratios $z:=\frac{(x_2-x_1)(x_4-x_3)}{(x_3-x_1)(x_4-x_2)}$ and $\bar{z}:=\frac{(\bar{x}_2-\bar{x}_1)(\bar{x}_4-\bar{x}_3)}{(\bar{x}_3-\bar{x}_1)(\bar{x}_4-\bar{x}_2)}$ in \eqref{cbe} are real value as $z=\bar{z}$. 

The conformal blocks in \eqref{cbe} are not easily computable objects in general. However, in the semiclassical limit, which is defined by
\begin{align}
mnc\to\infty, \;\;\;\;\epsilon:= \frac{6h}{mnc} \quad \text{and} \quad \epsilon_p:= \frac{6h_p}{mnc} \quad \text{fixed},\label{scl}
\end{align}
the Virasoro conformal block $\mathcal{F}$ is expected to be exponentiated \cite{Belavin:1984vu, Zamolodchikov1987}
\begin{align}
\log\left[\mathcal{F}(mnc, h, h_p, 1-z)\right]
\,\, \sim \,\, -\frac{mnc}{6} \,\, f(\epsilon, \epsilon_p, 1-z),\label{expcb}
\end{align}
by an analysis of the Liouville theory.  
The author of \cite{Hartman:2013mia}, using \eqref{scl} and \eqref{expcb}, argued that \eqref{cbe} in the 2d holographic CFTs for some finite range around $z=1$ can be approximated by the single conformal block in $t$-channel with the lowest conformal dimension $h_p = h_\text{\,low}$.

In our case \eqref{cbe}, OPE in \eqref{OPE} determines the lowest conformal dimension for $t$-channel:
\begin{align}
\begin{split}
h_\text{\,low} = h_{g_Bg_A^{-1}} = \frac{2c}{24}\left(n-\frac{1}{n}\right) \,, \quad
\epsilon_p = \epsilon_{\text{low}}:=\frac{6h_\text{\,low}}{mnc} \,, \label{ep}
\end{split}
\end{align}
where, we use \eqref{ht2}. This is because the exchange of the unit operator is forbidden unless $n=1$. Accordingly, in the large $c$ limit with $m$ and $n$ held fixed, one can confirm that $\epsilon_p=\epsilon_{\text{low}}:=\frac{6h_\text{\,low}}{mnc}$ satisfies the semiclassical limit \eqref{scl}.

Plugging \eqref{expcb} into the \eqref{cbe} with $h_p = h_\text{\,low}$ and $\epsilon_p = \epsilon_\text{\,low}$, we obtain the following
\begin{align}
&\log\left\langle\sigma_{g_A}(x_1)\sigma_{g_A^{-1}}(x_2)\sigma_{g_B}(x_3)\sigma_{g_B^{-1}}(x_4)\right\rangle_{CFT^{\otimes mn}}\notag\\
\sim&-4h\log[(x_4-x_1)(x_3-x_2)]+2\log C_{n, m}-\frac{mnc}{3}f(\epsilon, \epsilon_\text{low}, 1-z),\label{cf1}
\end{align}
where $C_{n, m}$ is the OPE coefficient $C_{ABp}$ with exchange of $\sigma_{g_Bg_A^{-1}}$. Its explicit form is given by \cite{Dutta:2019gen, Lunin:2000yv}
\begin{align}
C_{n, m}=(2m)^{-4h_n}, \;\;h_n=\frac{c}{24}\left(n-\frac{1}{n}\right).\label{cnm}
\end{align}
Using \eqref{cf1}, the denominator of \eqref{RRE} $\left\langle\sigma_{g_m}(x_1)\sigma_{g_m^{-1}}(x_2)\sigma_{g_m}(x_3)\sigma_{g_m^{-1}}(x_4)\right\rangle_{CFT^{\otimes m}}$ can be computed by
\begin{align}
&\log\left\langle\sigma_{g_m}(x_1)\sigma_{g_m^{-1}}(x_2)\sigma_{g_m}(x_3)\sigma_{g_m^{-1}}(x_4)\right\rangle_{CFT^{\otimes m}}\notag\\
=& \,\,\, \lim_{n\to1}\log\left\langle\sigma_{g_A}(x_1)\sigma_{g_A^{-1}}(x_2)\sigma_{g_B}(x_3)\sigma_{g_B^{-1}}(x_4)\right\rangle_{CFT^{\otimes mn}}\notag\\
\sim& \,\,\, \lim_{n\to1}\left[-4h\log[(x_4-x_1)(x_3-x_2)]+2\log C_{n, m}-\frac{mnc}{3}f(\epsilon, \epsilon_\text{low}, 1-z)\right],\label{cf2}
\end{align}
where, the first equality is justified because  \eqref{EQUAL} and \eqref{cf1} is used in the last line.

Since $\epsilon$ in \eqref{scl} and $\epsilon_\text{low}$ in \eqref{ep} are proportional to $m-1$ and $n-1$ respectively, $\epsilon$ and $\epsilon_\text{low}$ become small  around $m=1$ and $n=1$. Thus we can express \eqref{cf1} and \eqref{cf2} using a perturbative expansion about $\epsilon \ll 1$ and $\epsilon_\text{low} \ll 1$. The perturbative expansion of $f(\epsilon, \epsilon_\text{low}, 1-z)$ in $\epsilon$ and $\epsilon_\text{low}$ is given as \cite{Fitzpatrick:2014vua}\footnote{The formula (\ref{sccb}) for $t$-channel is obtained from the formula (D.24) for $s$-channel in \cite{Fitzpatrick:2014vua} with an exchange $z\leftrightarrow1-z$. Since our definition of the Virasoro conformal block $\mathcal{F}$ does not include $1/(x_3-x_2)^{2h}$ as shown in (\ref{cbe}), (\ref{sccb}) does not include $2\epsilon\log[1-z]$.}:
\begin{align}
f(\epsilon, \epsilon_\text{low}, 1-z)=&\epsilon_{\text{low}}\log\left[\frac{1+\sqrt{z}}{4(1-\sqrt{z})}\right]+(2\epsilon^2-\epsilon_\text{low}^2)\log z+2\epsilon_\text{low}^2\log\left[\frac{1}{2}(1+\sqrt{z})\right]\notag\\
+&(\epsilon_\text{low}-2\epsilon )^2+\frac{(\epsilon_\text{low}-2\sqrt{z}\epsilon )^2\log z}{1-z}+\cdots,\label{sccb}
\end{align}
where $\cdots$ means that we consider the perturbation up to quadratic order in $\epsilon$ and $\epsilon_\text{low}$.  

Finally, putting \eqref{cf1} with $C_{n, m}$ in \eqref{cnm} and $f(\epsilon, \epsilon_\text{low}, 1-z)$ in \eqref{sccb} into \eqref{RRE}, we obtain the reflected entropy in the 2d holographic CFTs up to first order in $m-1$:
\begin{align}
\lim_{n\to1}S_n(AA^\star)_{\psi_m}\sim\frac{c}{3}\log\left[\frac{1+\sqrt{z}}{1-\sqrt{z}}\right]-\frac{2c(m-1)}{3}\frac{\sqrt{z}\log z}{1-z}+\mathcal{O}((m-1)^2).\label{SNM}
\end{align}
The first term in (\ref{SNM}) is the reflected entropy for $\rho_{AB}$, which was computed in \cite{Dutta:2019gen}\footnote{The cross ratio $x$ in \cite{Dutta:2019gen} is related to our cross ratio $z$ as $x=1-z$. }, and the second term is the first order correction in $m-1$. Note that (\ref{SNM}) is valid for some finite range of $z$ around $z = 1$ because we use the conformal block in $t$-channel.
Let us sketch how the leading (and sub-leading) terms of \eqref{SNM} in the $m-1$ expansion are obtained. 
Note that the result \eqref{SNM} is originated from $n-1$ order contributions in \eqref{cf1} through out the formula \eqref{RRE}\footnote{The higher order contribution  $\mathcal{O}((n-1)^2)$ will vanish after taking $n\to1$ limit.}. Because of the following facts with the series expansion by $n-1$ and $m-1$,
\begin{align}
\log C_{n, m} \propto c_{1}(n-1) +  c_{2}(n-1)(m-1) \,, \quad \epsilon_{\text{low}} \propto n-1 \,, \quad  \epsilon \propto m-1 \,, \label{}
\end{align}
one can notice that there are three $n-1$ order terms in \eqref{cf1} using \eqref{sccb}\footnote{Strictly speaking, $\epsilon_{\text{low}}$ depends on $m$ and includes the sub-leading term of $(n-1)(m-1)$ order. However, because of $mnc$ factor for $mncf(\epsilon, \epsilon_\text{low}, 1-z)$ in (\ref{cf1}), the final result does not depend on this sub-leading term. Another logarithm term $4h\log[(x_4-x_1)(x_3-x_2)] $ also does not contribute to the final result due to the cancelation.}: i) $\log C_{n, m}$, ii) $\epsilon_\text{low}$ - order, iii) $\epsilon_\text{low}\,\epsilon$ - order. 
Then, we can finally see which contributions make the leading (and sub-leading) terms in \eqref{SNM} as follows
\begin{align}
\begin{split}
\text{Leading term:}& \quad  \log C_{n, m} \propto  c_{1}(n-1) \,, \,\,\qquad\qquad \epsilon_\text{low} \,\,\,\, \propto (n-1) \,, \\
\text{Sub-leading term:}& \quad  \log C_{n, m} \propto  c_{2}(n-1)(m-1) \,, \quad \epsilon_\text{low}\,\epsilon \propto (n-1)(m-1) \,.
\end{split}
\end{align}
%

%%%%%%%%%%%%%%%%%%%%%%%%%%%%%%%%%%%%%%%%%
%
%        Section: Entanglement wedge cross section with small backreaction
%
%%%%%%%%%%%%%%%%%%%%%%%%%%%%%%%%%%%%%%%%%

\section{Entanglement wedge cross section with the small backreaction}\label{sec3}
In this section, we compute the entanglement wedge cross section for two intervals at the boundary of AdS$_3$ with the small backreaction from cosmic branes which are anchored at boundaries of the intervals. In particular, we evaluate a first order correction in $m-1$ to the entanglement wedge cross section, where $m$ is related to the tension of the cosmic branes $T_m = \frac{m-1}{4mG_N}$ with the gravitational constant $G_{N}$. As the QFT dual, $m$ is carried out through the replica index of $\rho_{AB}^m$. When the replica index $m$ is 1, the cosmic branes become tensionless minimal surfaces, and they no longer backreact on the geometry, reproducing the Ryu-Takayanagi surface. Thus we can think of the cosmic branes as an extension of the Ryu-Takayanagi surface in $m\neq1$ direction.  
Adding one more description of  holographic setup, one might wonder what the holographic interpretation of the other replica index $n$ of CFTs is. It is, in the same way as the cosmic brane, related to the tension of the cosmic branes in the entanglement wedge~\cite{Kusuki:2019zsp}. Similarly to the CFTs in previous section, we focused on the perturbative expansion of $m$ only. Therefore, in this paper, we will consider the tensionless cosmic branes ($n=1$) in the entanglement wedge. As a methodological perspective, we apply the same prescription given in~\cite{Dong:2016fnf}, which is used to obtain the minimal area of cosmic branes anchored at the AdS boundary, to compute the entanglement wedge cross section up to first order in $m-1$. Then, we compare the entanglement wedge cross section to the reflected entropy \eqref{SNM} in the previous section and explicitly show the duality between them.

\subsection{Entanglement wedge cross section: a quick review}
\paragraph{Entanglement wedge cross section $E_W(A: B)$ \textit{without} backreaction:}
%We start explaining, without considering backreaction, the minimal surfaces of two intervals ($A, B$) at the AdS boundary with fixing a time slice $t=0$: $A=[x_1, x_2]$ and $B=[x_3, x_4]$ with $x_1<x_2<x_3<x_4$. 
We start explaining, without considering the backreaction, the minimal surfaces of two intervals in the pure AdS$_3$:
\begin{align}
\dd s^2 = \frac{\ell^2}{\xi^2} \dd \xi^2 + \frac{\xi^2}{\ell^2}\left(\dd t^2 +\dd x^2  \right), \label{PM}
\end{align}
where the AdS boundary is located at $\xi \rightarrow \infty$, and $\ell$ is the AdS radius which will be set to one for simplicity. Two intervals ($A, B$) of our interest are placed at the AdS boundary at a fixed time slice $t=0$: $A=[x_1, x_2]$ and $B=[x_3, x_4]$ with $x_1<x_2<x_3<x_4$.

\begin{figure}[t]
\centering
     \subfigure[Disconnected minimal surface]
     {\includegraphics[width=6.3cm]{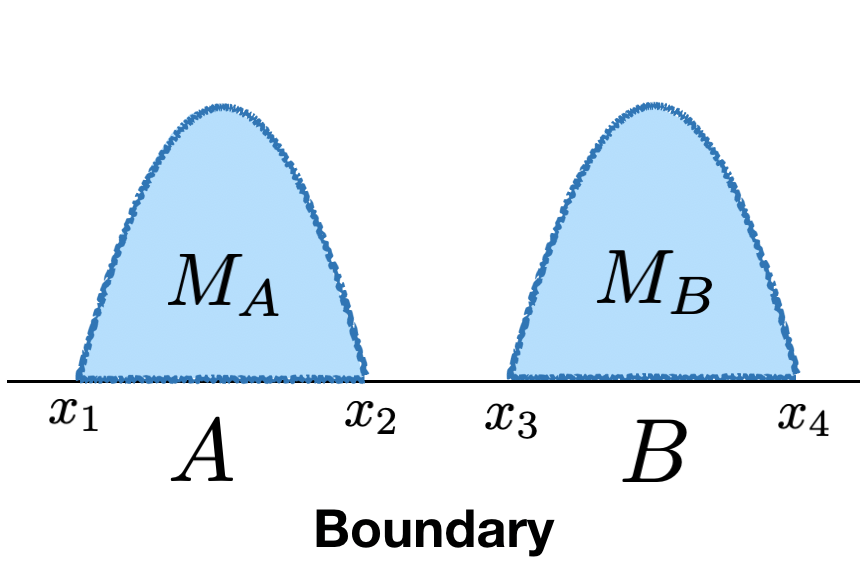} \label{DF}}\hspace{1cm}
      \subfigure[Connected minimal surface]
     {\includegraphics[width=6.3cm]{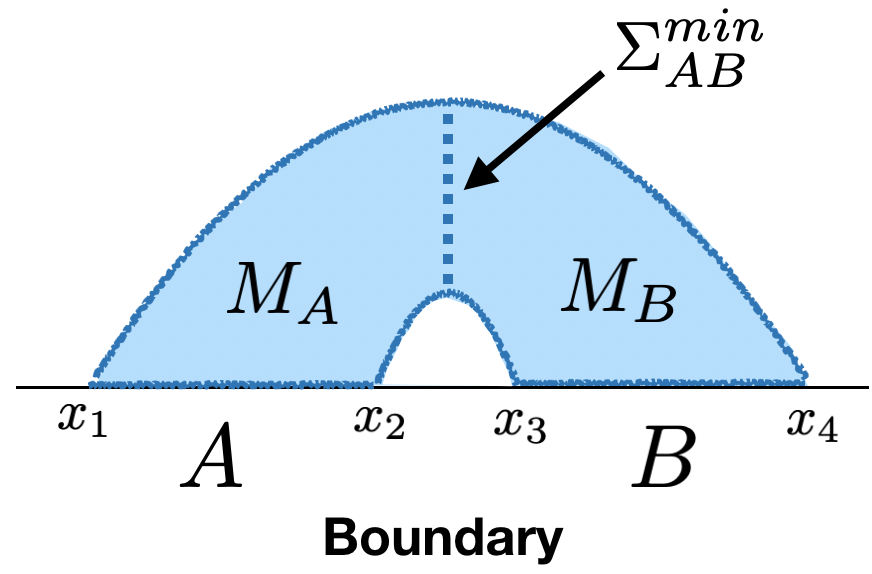} \label{CF}}
 \caption{Schematic pictures of the minimal surface $\Gamma_{AB}^{min}$ (the blue curves) and the entanglement wedge $M_{AB}$ (the blue shaded region). The blue dashed line in (b) represents the minimal surface  $\Sigma_{AB}^{min}$ in  $M_{AB}$, and it plays a role of dividing $M_{AB}$ into $M_A$ and $M_B$.}\label{MS}
\end{figure}

In this set-up, we have two possible configurations of the minimal surfaces $\Gamma_{AB}^{min}$ for $A \cup B$. One is a disconnected minimal surface (Fig. \ref{DF}), and the other is a connected minimal surface (Fig. \ref{CF}). The question to ask here is which configuration is the dominant minimal surface. The answer to this question depends on the cross ratio $z:=\frac{(x_2-x_1)(x_4-x_3)}{(x_3-x_1)(x_4-x_2)}$. The disconnected surface is dominant in $0<z<1/2$, whereas the connected surface is dominant in $1/2< z<1$~\cite{Headrick:2010zt}.

Next, we define the entanglement wedge cross section $E_W(A: B)$ based on entanglement wedge $M_{AB}$~\cite{Takayanagi:2017knl, Nguyen:2017yqw}. The entanglement wedge $M_{AB}$ (the blue shaded region in Figure \ref{MS}) is defined by a region whose boundary is $\partial M_{AB}=A\cup B\cup \Gamma_{AB}^{min}$. 
Inside the entanglement wedge $M_{AB}$, we can consider the minimal surface $\Sigma_{AB}^{min}$ which divides $M_{AB}$ into $M_A$ and $M_B$ where $\partial M_{A}\supset A$ and $\partial M_{B}\supset B$. This $\Sigma_{AB}^{min}$ is displayed as a blue dashed line in Fig. \ref{CF}.
Using the area of $\Sigma_{AB}^{min}$, we can finally define the entanglement wedge cross section $E_W(A: B)$ as
\begin{align}
E_W(A: B):=\frac{\text{Area}[\Sigma_{AB}^{min}]}{4G_N} \,, \label{EW} 
\end{align}
where $G_{N}$ is the gravitational constant.
Note that $E_W(A: B)=0$ for the disconnected surface since $M_{AB}$ for the disconnected minimal surface is initially disconnected ($\text{Area}[\Sigma_{AB}^{min}]=0$).
%Since $M_{AB}$ with the disconnected minimal surface $\Gamma_{AB}^{min}$ is initially disconnected, $E_W(A: B)$ for the disconnected surface $\Gamma_{AB}^{min}$ is zero, and there is a phase transition of $E_W(A: B)$ at $z=1/2$.

\paragraph{Entanglement wedge cross section $E_W(A: B)$ \textit{with} backreaction:}
We will shortly explain how the backreacted geometry can be introduced. Before doing so, we first give the reformed entanglement wedge cross section formula by a backreaction of cosmic brane:
\begin{align}
E_{mW}(A: B):=\frac{\text{Area}[\Sigma_{mAB}^{min}]}{4G_N}.\label{EmW}
\end{align}
Note that equation \eqref{EmW} has one more index $m$ than \eqref{EW}. This $m$ represents the replica index in the field theory and is related to the tension $T_m$ of the cosmic branes in the gravity theory via $T_m = \frac{m-1}{4mG_N}$~\cite{Dong:2016fnf}.
This reformulated entanglement wedge cross section \eqref{EmW} is obtained by replacing $\Sigma_{AB}^{min}$ in \eqref{EW} with the backreacted minimal surface $\Sigma_{mAB}^{min}$, in other words, the minimal surface $\Gamma_{AB}^{min}$ is replaced by the cosmic branes giving the conical singularity with the tension~\cite{Vilenkin:1981zs}.
%If we replace the minimal surface $\Gamma_{AB}^{min}$ by the cosmic branes with the tension $T_m$, the cosmic branes make backreaction to the geometry (\ref{PM}) and a conical opening angle $2\pi/m$ \cite{Vilenkin:1981zs, Dong:2016fnf}. Because of this backreaction, the minimal surface $\Gamma_{mAB}^{min}$ of cosmic branes is deformed from $\Gamma_{AB}^{min}$, where we use an index $m$ to express the backreaction. By using this minimal surface $\Gamma_{mAB}^{min}$, we can also define the entanglement wedge $M_{mAB}$ and the minimal surface $\Sigma_{mAB}^{min}$ in $M_{mAB}$. Then, the entanglement wedge cross section $E_{mW}(A: B)$ with the backreaction is defined by
%\begin{align}
%E_{mW}(A: B):=\frac{\text{Area}[\Sigma_{mAB}^{min}]}{4G_N}.\label{}
%\end{align}
%With $m=1$, (\ref{EmW}) reduces to (\ref{EW}) in the geometry (\ref{PM}).

{Generally, for the computation of $E_W(A: B)$ with the two intervals $A$ and $B$, we need to consider the backreaction from the two cosmic branes together. However, at the first order in $m-1$, we do not need to consider the two backreaction together because the simultaneous backreaction from the two cosmic branes is only affected by the second and higher order. Therefore, $E_W(A: B)$ at the first order in $m-1$ can be computed by the sum of $E_W(A: B)$ with the backreaction from the single cosmic brane. From the next subsection, we will compute $E_W(A: B)$ with the backreaction from the single cosmic brane.}

\subsection{Explicit computation of $E_{mW}(A: B)$ up to first order in $m-1$}
The 3d bulk geometry for Einstein gravity with the backreaction from the single cosmic brane can be described by~\cite{Dong:2016fnf, Hung:2011nu}
\begin{align}
\dd s^2=\frac{\dd r^2}{r^2-r_{h}^2}+(r^2-r_{h}^2)\dd \tau^2+r^2\dd \rho^2 \,, \label{BHM}
\end{align}
where we have the black hole horizon as $r_h = 1/m$, and the period of $\tau$ is fixed as $2\pi$. Here, the cosmic brane covers the horizon and is anchored at $\rho=-\infty$ and $\rho=\infty$. The reason why this metric \eqref{BHM} is  used to express the bulk geometry with the cosmic brane is that \eqref{BHM} includes  the same conical singularity of the cosmic brane at the horizon. Let us see the near horizon geometry of \eqref{BHM} as,
\begin{align}
\begin{split}
\dd s^2 |_{r\sim r_h}  \,\sim\, & \frac{\dd r^2}{\frac{2}{m}(r-r_h)}+\frac{2}{m}(r-r_h)\dd \tau^2+r_h^2\dd\rho^2 \\
\,=\, &\dd\tilde{r}^2+\tilde{r}^2\dd\left(\frac{\tau}{m}\right)^2+r_h^2\dd\rho^2,\label{MnH}
\end{split}
\end{align}
where $\tilde{r}:=\sqrt{2m(r-r_h)}$. When we fix the period of $\tau$ as $2\pi$, the metric (\ref{MnH}) has a conical opening angle $2\pi/m$ at $r\sim r_h$. In addition to the view of conical singularity from cosmic brane, there is another way to see this conical singularity in other language: the quotient replica manifold $\mathcal{B}_m/\mathbb{Z}_m$~\cite{Lewkowycz:2013nqa,  Faulkner:2013yia}\footnote{The main logic of it is as follows. We can think of the periodicity around a fixed point on the bulk replica manifold $\mathcal{B}_m$ as $2\pi m$. Then, by taking a quotient by $\mathbb{Z}_m$ replica symmetry, this periodicity is changing into $2\pi$ with the conical singularity therein. These $2\pi$ periodicity and conical singularity are related to the periodicity of $\tau$ in \eqref{BHM} and the singularity in \eqref{MnH}, respectively. For a comprehensive review of this, see \cite{Rangamani:2016dms}, for example.}. 
%Fixing the periodicity $\theta$ with $2\pi m$ around a fixed point on the bulk replica manifold $\mathcal{B}_m$, the periodicity $\theta$ is changed on the quotient $\mathcal{B}_m/\mathbb{Z}_m$ as $2\pi$ with the conical singularity therein. Therefore, due to the same periodicity $2\pi$ appearing in $\mathcal{B}_m/\mathbb{Z}_m$ and \eqref{BHM} with the conical singularity, the geometry \eqref{BHM} can be considered as the quotient $\mathcal{B}_m/\mathbb{Z}_m$ of the bulk spacetime, which we desire for the holographic construction of bulk replica manifold.
%Using GKP-Witten formula, we may write the partition function $Z$ of the boundary theory with on-shell action of the bulk solution $I_{\text{bulk}}$ as
%
%\begin{align}
%\begin{split}
%Z = e^{-I_{\text{bulk}}\left[\mathcal{B}_m\right]} \,
%\end{split}
%\end{align}
%
%where $\mathcal{B}_m$ is the bulk replica manifold. Since the bulk action is local, we may write
%
%\begin{align}
%\begin{split}
%I_{\text{bulk}}\left[\mathcal{B}_m\right] = m \, I_{\text{bulk}} \, [\hat{\mathcal{B}}_m]\,, \qquad \hat{\mathcal{B}}_m := \mathcal{B}_m/\mathbb{Z}_m \,,
%\end{split}
%\end{align}
%
%where, $\hat{\mathcal{B}}_m$ is defined by taking a quotient by $\mathbb{Z}_m$ replica symmetry on $\mathcal{B}_m$.

Let us explain how coordinates of backreacted geometry $(r, \tau, \rho)$ in \eqref{BHM} can be related to the coordinate of two intervals $(\xi, t, x)$ in \eqref{PM} by following the same strategy in~\cite{Dong:2016fnf}\footnote{In the appendix of \cite{Dong:2016fnf}, the bulk geometry for the disconnected minimal surface was considered. Thus, our coordinate transformation is different from one in \cite{Dong:2016fnf}.}.
By using an appropriate conformal transformation on \eqref{PM}, we can start with:
\begin{align}
x_1\to-1, \;\;x_2\to-R_0,\;\; x_3\to R_0,\;\; x_4\to 1, \label{TR1}
\end{align}
where $0<R_0<1$. Since the cross ratio $z := \frac{(x_1-x_2)(x_3-x_4)}{(x_1-x_3)(x_2-x_4)}$ is invariant under a global conformal transformation, $R_0$ is determined by 
\begin{align}
z=\frac{(1-R_0)^2}{(1+R_0)^2}. \label{CROSS}
\end{align}
In addition to the transformation \eqref{TR1}, we use a following conformally map
\begin{align}
\tan{\tau}=\frac{2t}{1-t^2-x^2},\;\; \tanh{\rho}=\frac{2x}{1+t^2+x^2}, \label{CONFORMAL}
\end{align}
where the period of $\tau$ is $2\pi$. Then, the intervals are conformally mapped as
\begin{align}
\begin{split}
&A:\quad -1 \le \,x \,\le -R_{0}    \,\,\,\,\quad (t=0)     \quad\rightarrow\quad     -\infty\le\rho\le-\rho_{0}         \quad(\tau=0) \,, \\
&B:\quad R_{0} \le \,x \, \le 1    \,\qquad\quad (t=0)     \quad\rightarrow\quad\,\,\,\,\,  \rho_{0}\le\rho\le\infty   \,\,\,\,\,\quad(\tau=0) \,, \label{TR2}
\end{split}
\end{align}
where
\begin{align}
\rho_0:=\text{arctanh}\frac{2R_0}{1+R_0^2}=-\frac{1}{2}\log{z}. \label{RHO0}
\end{align}
In the last equality in \eqref{RHO0}, the cross ratio \eqref{CROSS} is used. 
\begin{figure}[]
\centering
     {\includegraphics[width=15cm]{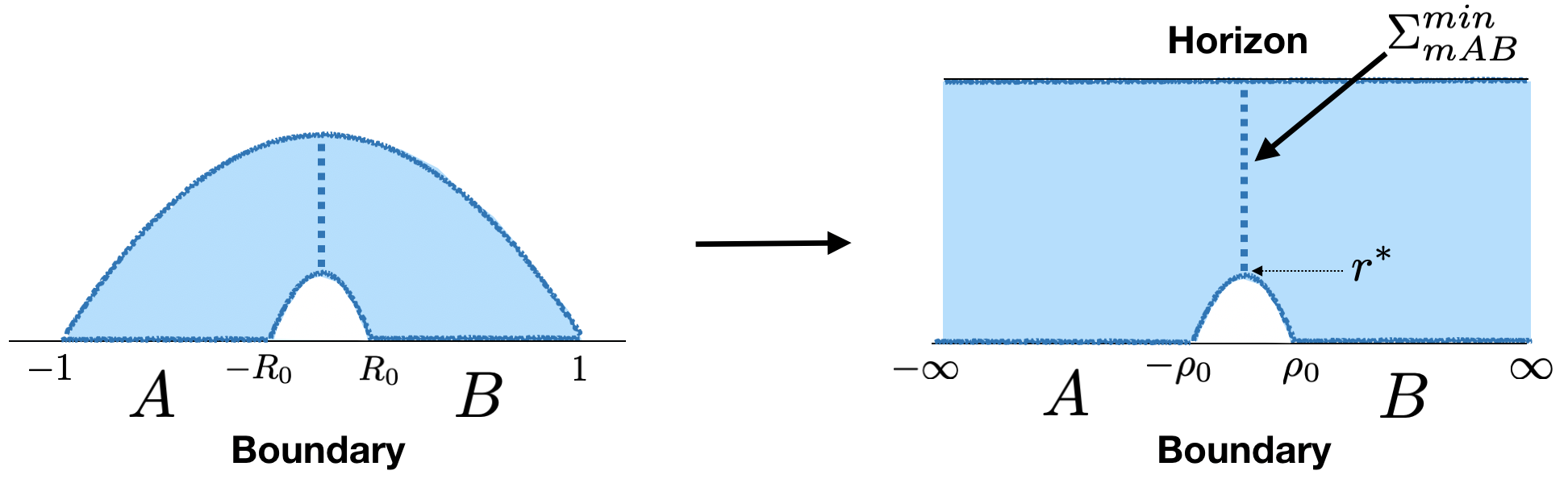} \label{}}
 \caption{The change of configuration of two intervals by the conformal transformation \eqref{TR2}. The blue dashed line is the minimal surface $\Sigma_{mAB}^{min}$, $\rho_{0}$ is given by \eqref{RHO0}, and $r^{*}$ is determined from \eqref{rSTAR}.}\label{EWBH}
\end{figure} 
The change of configuration of two intervals along \eqref{TR2} are displayed in Fig. \ref{EWBH}.
Furthermore, under the conformal transformation \eqref{CONFORMAL}, the 2d flat metric at the AdS boundary in \eqref{PM}, $\dd s^2_{\mathbb{R}^2}=\dd t^2+\dd x^2$, is mapped to 
\begin{align}
\dd s^2_{S^1\times\mathbb{R}}=\dd \tau^2+\dd \rho^2,\label{dsSR}
\end{align} 
up to the pre-factor.
According to the fact that the metric given in \eqref{dsSR} is conformally equivalent to \eqref{BHM} at the boundary, we can use the backreacted geometry \eqref{BHM} to compute the area of $\Sigma_{mAB}^{min}$  for the two intervals $A$ and $B$.

Next, we genuinely compute the area of the minimal surface $\Sigma_{mAB}^{min}$ in the geometry \eqref{BHM}, which includes the backreaction from the \textit{single} cosmic brane. As shown in Fig. \ref{EWBH}, $\Sigma_{mAB}^{min}$ is placed between $r = r_h$ and $r = r_{*}$ at $\rho=0$. Here $r_{*}$ is determined as a value of $r$ on minimal surface $\Gamma_{mAB}^{min}$ at $\rho=0$, which is placed in $-\rho_0<\rho<\rho_0$, and is given as~\cite{Faraggi:2007fu, Rangamani:2016dms, Kudler-Flam:2018qjo}
\begin{align}
r_* = r_h\coth[r_h \, \rho_0] = \frac{1}{m}\coth\left[\frac{\rho_0}{m}\right], \label{rSTAR}
\end{align}
where we use $r_h = 1/m$ in the last equality.
Then, using the area formula, we can directly compute the area of $\Sigma_{mAB}^{min}$:
\begin{align}
\text{Area}[\Sigma_{mAB}^{min}]=&\int_{r_h}^{r_*}\frac{\text{d}r}{\sqrt{r^2-r_h^2}}=\log\left[\coth\frac{\rho_0}{2m}\right]\notag\\
=&\log\left[\frac{1+\sqrt{z}}{1-\sqrt{z}}\right]-(m-1)\frac{\sqrt{z}\log{z}}{1-z}+\mathcal{O}\left((m-1)^2\right).\label{AMS}
\end{align}
Here, we replaced $\rho_{0}$ with the cross ratio $z$ via \eqref{RHO0}.
The final result of  $\text{Area}[\Sigma_{mAB}^{min}]$ in \eqref{AMS} consists of two terms. The first term corresponds to the minimal area without the backreaction, and the second term shows the first order correction in ($m-1$) from the \textit{single} cosmic brane.
%\begin{figure}[t]
%\centering
%     {\includegraphics[width=6cm]{EWBH} \label{EWBH}}
% \caption{Schematic pictures of the minimal surface $\Sigma_{mAB}^{min}$ (the black dashed line) in the geometry (\ref{BHM}).    }
%\end{figure} 

To complete the full calculation of the area of the minimal surface $\Sigma_{mAB}^{min}$ of the \textit{two} cosmic branes, we also need to consider the contribution from the other cosmic brane anchored at $\rho=-\rho_{0}$ and $\rho=\rho_{0}$. It can be done by considering a transformation on \eqref{PM}:
\begin{align}
x_1\to-R_0, \;\;x_2\to-1,\;\; x_3\to 1,\;\; x_4\to R_0 \,. \label{TR3}
\end{align}
After using this transformation, one can notice that the cosmic brane anchored at $\rho=-\rho_{0}$ and $\rho=\rho_{0}$ with the transofmration \eqref{TR1} is now located at $\rho=-\infty$ and $\rho=\infty$ with \eqref{TR3}. Thus we can apply the same procedure used in the previous paragraph, and we will have the same result as \eqref{AMS}.

Using the definition given in \eqref{EmW}, we can summarize that the entanglement wedge cross section $E_{mW}(A: B)$ of the connected minimal surface with the backreaction from the two cosmic branes is
\begin{align}
E_{mW}(A: B)=\frac{1}{4G_N}\log\left[\frac{1+\sqrt{z}}{1-\sqrt{z}}\right]-\frac{(m-1)}{2G_N}\frac{\sqrt{z}\log{z}}{1-z}+\mathcal{O}\left((m-1)^2\right).\label{EmW2}
\end{align}
Note that when the replica index $m$ approaches to 1, \eqref{EmW2} reproduces the entanglement wedge cross section $E_{W}(A: B)$ without the backreaction~\cite{Takayanagi:2017knl}\footnote{Our definition of the cross ratio $z$  is different from one in \cite{Takayanagi:2017knl}. By replacing the cross ratio in (\ref{EmW2}) with $m=1$ as $z\to z/(z+1)$, one can obtain the expression in \cite{Takayanagi:2017knl}.}.

As a main result of this paper, now we can show that, even in the presence of the backreaction from the cosmic branes, the holographic calculation \eqref{EmW2} perfectly matches with the field theory calculation in \eqref{SNM}:
%
%\begin{align}
%\lim_{n\to1}S_n(AA^\star)_{\psi_m} \,\sim\, 2E_{mW}(A: B)+\mathcal{O}((m-1)^2),
%\end{align}
\begin{align}
\begin{split}
2E_{mW}(A: B) &= \frac{1}{2G_N}\log\left[\frac{1+\sqrt{z}}{1-\sqrt{z}}\right]-\frac{(m-1)}{G_N}\frac{\sqrt{z}\log{z}}{1-z}+\mathcal{O}\left((m-1)^2\right) \, \\
&= \frac{c}{3}\log\left[\frac{1+\sqrt{z}}{1-\sqrt{z}}\right]-\frac{2c(m-1)}{3}\frac{\sqrt{z}\log z}{1-z}+\mathcal{O}((m-1)^2)
\end{split}
\end{align}
where we used $c=\frac{3}{2G_N}$~\cite{Brown:1986nw} in the last equality.
%where $\sim$ denotes that we suppress sub-leading terms in a large central charge limit $\mathcal{O}((1/c)^0)$.
This is an explicit check of the duality (\ref{d2}) between the reflected entropy and the entanglement wedge cross section without the quantum correction up to first order in $m-1$.

%%%%%%%%%%%%%%%%%%%%%%%%%%%%%%%%%%%%%%%%%
%
%        Section: Summary and discussion
%
%%%%%%%%%%%%%%%%%%%%%%%%%%%%%%%%%%%%%%%%%
\section{Summary and discussion}\label{sec4}
In this paper, we have studied the following holographic duality giving a surprising relationship between the reflected entropy and the entanglement wedge cross section \cite{Dutta:2019gen}:
\begin{align}
\lim_{n\to1}S_n(AA^\star)_{\psi_m}=2E_{mW}(A: B)+\mathcal{O}(G_N^0) \,,\label{}
\end{align}
where, $\lim_{n\to1}S_n(AA^\star)_{\psi_m}$ is the reflected entropy for $\rho_{AB}^m$,  and $E_{mW}(A: B)$ is the entanglement wedge cross section in the quotient spacetime $\mathcal{B}_m/\mathbb{Z}_m$.
The main result of this paper is that we explicitly show this duality up to the first order in $m-1$. 
In the conformal field theory framework (CFT$_{2}$), $S_n(AA^\star)_{\psi_m}$ of two intervals $A$ and $B$ is expressed in terms of twist operators \eqref{RRE}. In the 2d holographic CFTs, we can compute $S_n(AA^\star)_{\psi_m}$ by using a perturbative expansion on the conformal block in the semiclassical limit (\ref{scl}) as shown in \eqref{sccb}. The final form of the reflected entropy from this field theory calculation is given in \eqref{SNM}.

On the other hand, in the gravity theory framework (AdS$_{3}$), the entanglement wedge cross section is computed in a backreacted bulk spacetime generated from cosmic branes. We used the fact that the pure AdS$_{3}$ \eqref{PM} with the backreaction from a single cosmic brane can be mapped to the backreacted black hole geometry \eqref{BHM} after doing several transformations \cite{Hung:2011nu}. Then, the entanglement wedge cross section is obtained to be the form as \eqref{EmW2} with the first order correction in $m-1$. By comparing the two main results from CFTs \eqref{SNM} and AdS$_{3}$ \eqref{EmW2}, we show that the holographic duality in \eqref{d2} is perfectly satisfied using $c=\frac{3}{2G_N}$.

%We have studied the duality (\ref{d2}) between the reflected entropy and the entanglement wedge cross section and explicitly shown it with the two disjoint intervals $A$ and $B$ up to first order in $m-1$. In the CFT$_2$ side, a perturbative expansion of the dominant conformal block in the 2d holographic CFTs determines the leading behavior of the reflected entropy. In the AdS$_3$ side, the cosmic branes  in the bulk geometry, which can be mapped to a black hole horizon, backreact the entanglement wedge cross section. Their corrections at first order in $m-1$ exactly satisfy the duality (\ref{d2}).  

We end with a description of some future works of interests.
One of the future directions from this study is a checking the duality at higher order terms in $m-1$. The monodromy method \cite{Hartman:2013mia, Fitzpatrick:2014vua, Harlow:2011ny} and the Zamolodchikov's recursion relation \cite{Zamolodchikov:1985ie, Zamolodchikov1987} might be useful to evaluate the dominant conformal block in the reflected entropy. For the entanglement wedge cross section at higher order in $m-1$, we need to consider the backreaction from two cosmic branes simultaneously, and it may be difficult to construct an analytic solution of the geometry. However, as used in section \ref{sec3}, the geometry with the backreaction from a single cosmic brane is known analytically \cite{Hung:2011nu}, and it is interesting to compare the entanglement wedge cross section in this geometry with some higher order terms in the conformal block.

Another future work is generalization to higher dimensional AdS/CFT. Since the computation method in  \cite{Dong:2016fnf} can be applied to the holographic R\'{e}nyi entropy between two disks in general dimensions, the entanglement wedge cross section in general dimensions may be also computable. On the other hand, we cannot use 2d CFT techniques to obtain the reflected entropy in general dimensions, so it is necessary to develop a procedure for an explicit computation. 
We leave these for future works.

\acknowledgments
We would like to thank Yuya Kusuki and Kotaro Tamaoka for discussions and comments. This work was supported by Basic Science Research Program through the National Research Foundation of Korea(NRF) funded by the Ministry of Science, ICT $\&$ Future Planning(NRF- 2017R1A2B4004810) and GIST Research Institute(GRI) grant funded by the GIST in 2019.  We also would like to thank ``Strings and Fields 2019'' in Kyoto, Japan, and the APCTP(Asia-Pacific Center for Theoretical Physics) focus program,``Quantum Matter from the Entanglement and Holography'' in Pohang, Korea for the hospitality during our visit, where part of this work was done.

%\bibliography{HyunSikRefs}
\bibliographystyle{JHEP}

\providecommand{\href}[2]{#2}\begingroup\raggedright\endgroup

\end{document}